\documentclass{aa}

\usepackage{graphicx}
\usepackage{txfonts}
\usepackage{caption}
\usepackage{subcaption}
\usepackage{tablefootnote}
\usepackage{hyperref}

\usepackage{comment}
\usepackage{soul}

\usepackage{color}
\usepackage{xcolor}

\begin{document}

\title{Global simulations of accretion flows onto perturbers embedded in magnetized 
disks -I. MRI and jet formation in ideal MHD}
   \titlerunning{Global simulations of accretion flows onto perturbers}

   \author{Raúl O. Chametla 
          \inst{1}
          \and 
          F. J. Sánchez-Salcedo\inst{2}
          \and 
          Martin E. Pessah\inst{3}
          \and
          Mauricio Reyes-Ruiz\inst{4}
          }
   \authorrunning{R.~O.~Chametla et al.}
   \institute{%
            Charles University, Faculty of Mathematics and Physics, Astronomical Institute,
            V Hole\v{s}ovi\v{c}k\'ach 747/2, 180 00,
            Prague 8, Czech Republic\\
            \email{raul@sirrah.troja.mff.cuni.cz}
             \label{UKarlova}
        \and
            Instituto de Astronomía, Universidad Nacional Autónoma de México, 
            Apt.~Postal 70-264, C.P.~04510, Mexico City, Mexico
             \label{UNAM}
         \and
            Niels Bohr International Academy, Niels Bohr Institute, Blegdamsvej 17, DK-2100 Copenhagen Ø, Denmark\\
             \label{NBIA}
        \and
            Instituto de Astronomía, Universidad Nacional Autónoma de México, Ensenada, 22800 B.C., México\label{UNAMIAE}
             }

   \date{Received XXX; accepted YYY}
   
  \abstract 
   {We present the highest resolution global MHD simulations to date of  gas flow around a low mass a perturber with mass ratio $q\in[10^{-4},10^{-3}]$, embedded in an accretion disk around a massive central object. We find that gas flow onto the secondary self-consistently forms a turbulent, magnetized mini-accretion disk. The mini-accretion disk sustains a large-scale magnetic field generated by the dynamo effect of the MRI and the accretion flow into the perturber. Simultaneously, a bipolar, collimated, magnetized outflow is launched, extending beyond the perturber's Hill sphere. The bipolar outflows are driven by the combined action of magnetic pressure, in the innermost regions of the mini-accretion disk, and the magnetocentrifugal acceleration of gas, which may attain speeds comparable to the escape velocity from the massive central object. Our results establish an important conceptual connection in accretion disk physics across a wide range of astrophysical systems—from mini-accretion disks to circumstellar and black hole accretion disks—by demonstrating that no fine-tuning is required for small-scale disks to naturally enter an outflow-launching regime. Beyond identifying the physical mechanism responsible for launching small-scale outflows, our framework lays the groundwork for developing more sophisticated physical models of mini-accretion disks around embedded low-mass perturbers.}


   \maketitle
%

\section{Introduction}
\label{sec:introduction}

\begin{figure*}
    \centering
    \includegraphics[width=1.0\linewidth]{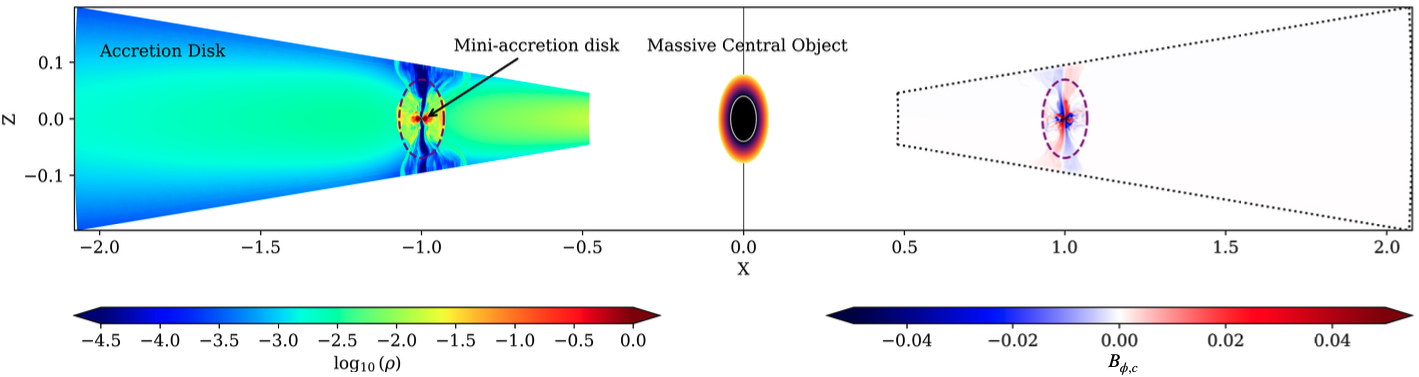}
    \caption{2D-maps in the $x-z$ plane of the gas density (left panel) and azimuthal component of the magnetic field (right panel) at $t=16.77\,T_0$ for $q=10^{-3}$. The dashed circle represents the Hill sphere.}
    \label{fig:global}
\end{figure*}

\begin{figure*}
    \centering
    \includegraphics[width=1.0\linewidth]{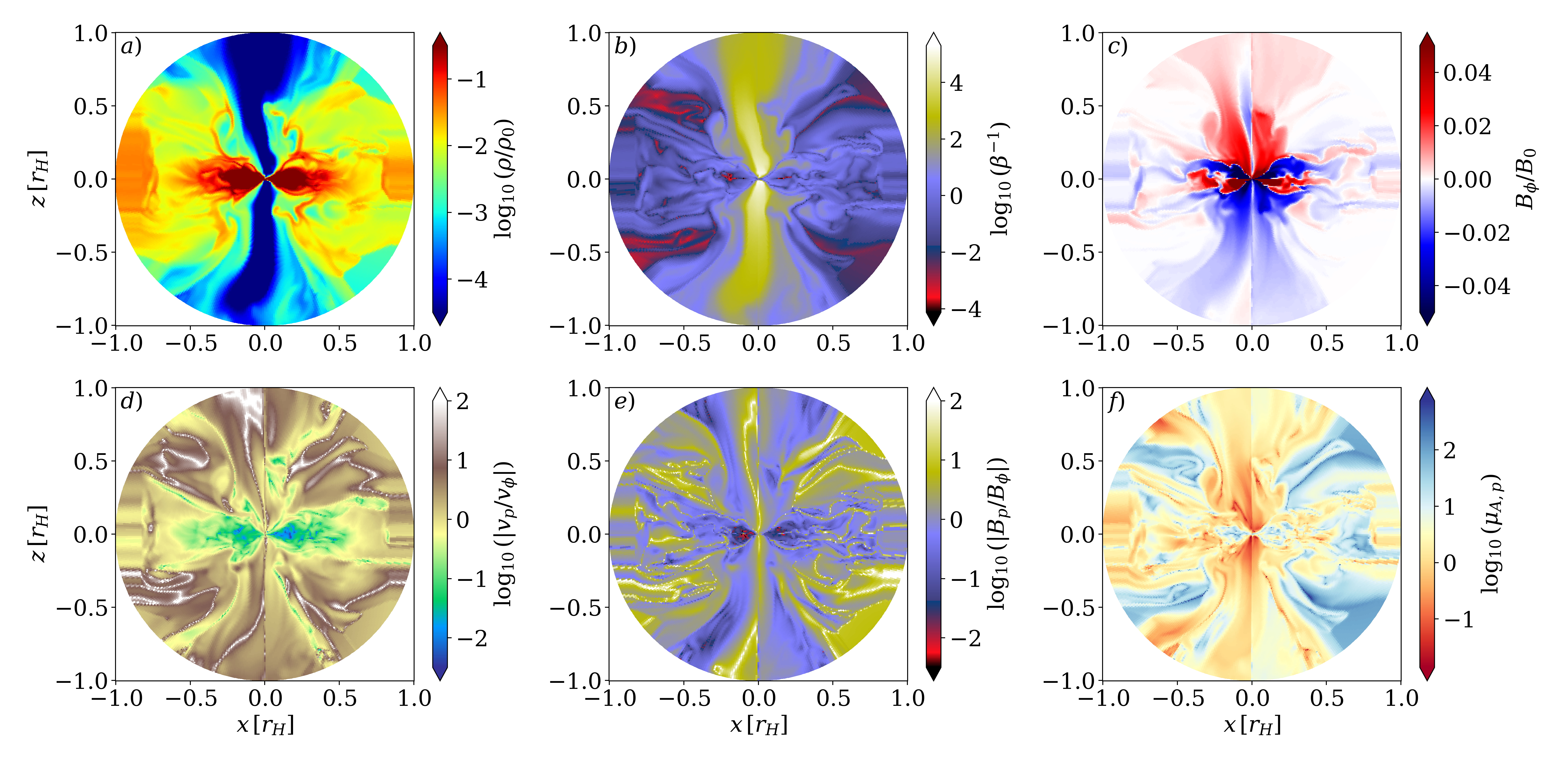}
    \caption{Vertical slices of the perturber's Hill sphere with $q=10^{-3}$ at $t=16.77\,T_0$ in the $x-z$ plane, perpendicular to the direction of motion of the perturber. The various panels illustrate the gas density, $\rho$, the inverse of the beta parameter, $\beta^{-1}$, the toroidal component of the magnetic field, $B_{\phi}$, the ratio of poloidal to toroidal components of the velocity, $|v_{\rm p}/v_{\phi}|$, the ratio of poloidal to toroidal components of the magnetic field, $|B_{\rm p}/B_{\phi}|$, and the poloidal Alfv\'en Mach number. The fiducial density, $\rho_0$, is related to the surface disk density via $\rho_0=\Sigma_0/(\sqrt{2\pi}H_{\rm p})$.
    }
    \label{fig:fields}
\end{figure*}

\begin{figure}
    \centering
    \includegraphics[width=1.0\linewidth]{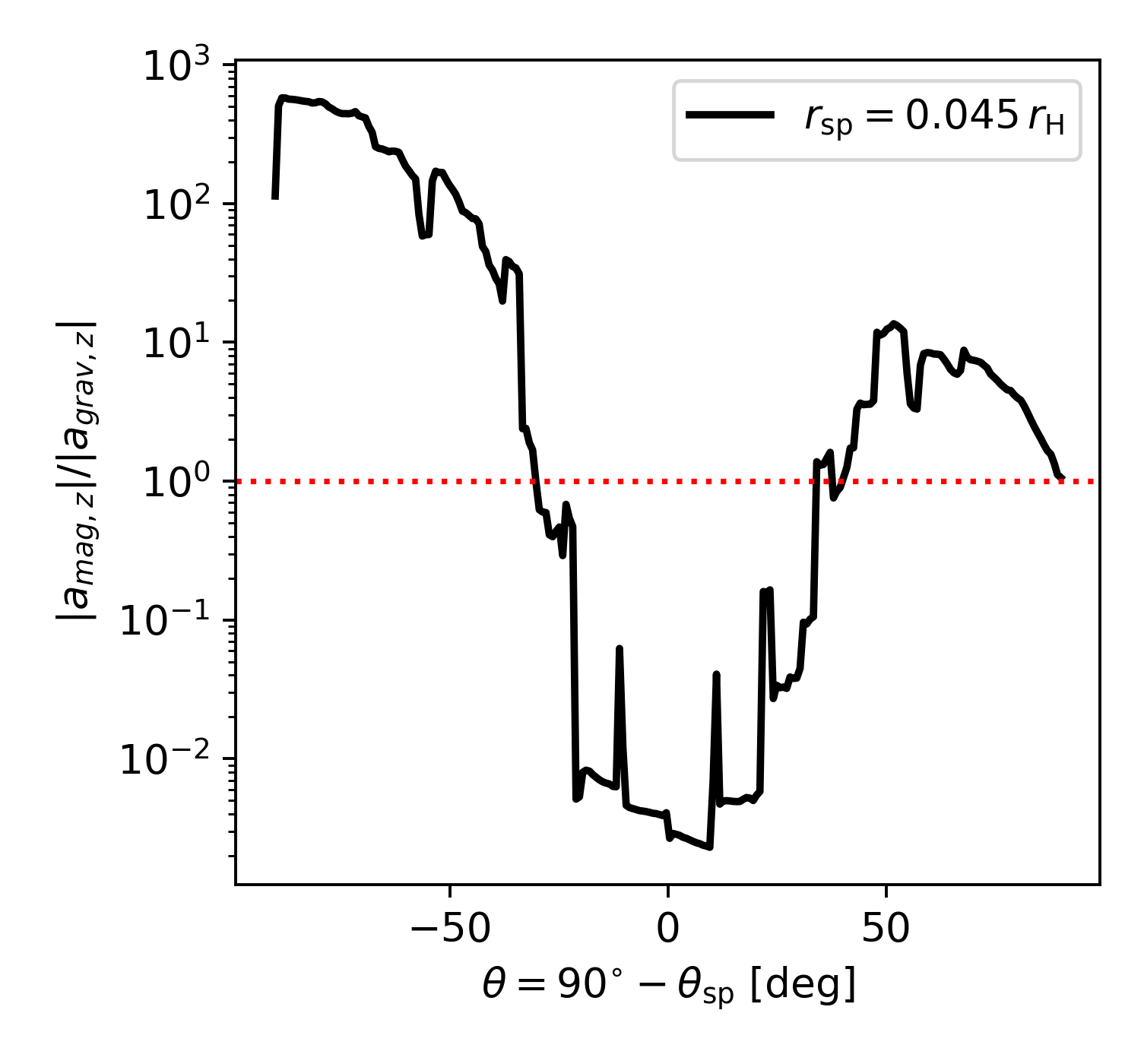}
    \caption{Magnetic-to-gravitational acceleration ratio in the vertical direction as a function of
    the latitude angle, at a fixed distance from the perturber ($r_{\mathrm{sp}}=0.045r_{H}$) for the $q=10^{-3}$ model. The vertical magnetic acceleration 
    is computed from the time-averaged magnetic pressure gradient (between $t=16.77-20.12\,T_0$).} 
    \label{fig:gradPm}
\end{figure}

\begin{figure*}
    \centering
    \begin{subfigure}[b]{0.5\textwidth} 
        \includegraphics[width=0.8\linewidth]{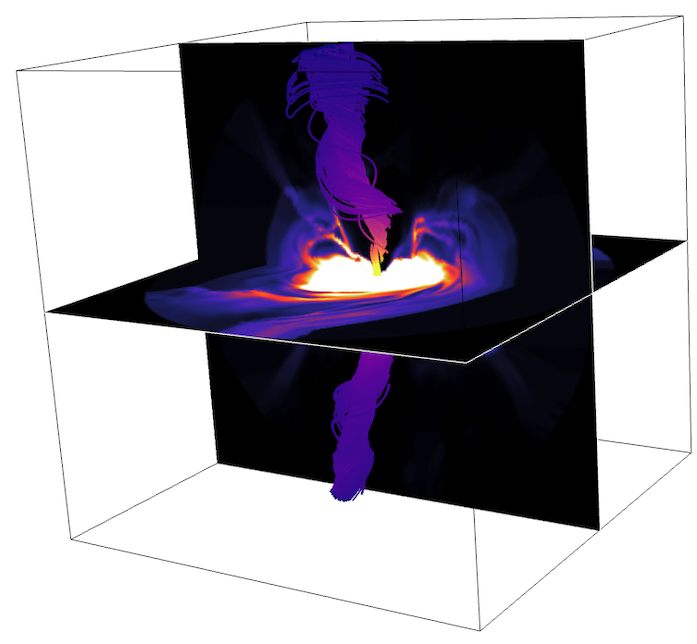}
        \caption{Magnetic field lines in the vicinity of the perturber.
        }
        \label{fig:sub1}
    \end{subfigure}%
    \hfill 
    \begin{subfigure}[b]{0.5\textwidth}
        \includegraphics[width=\linewidth]{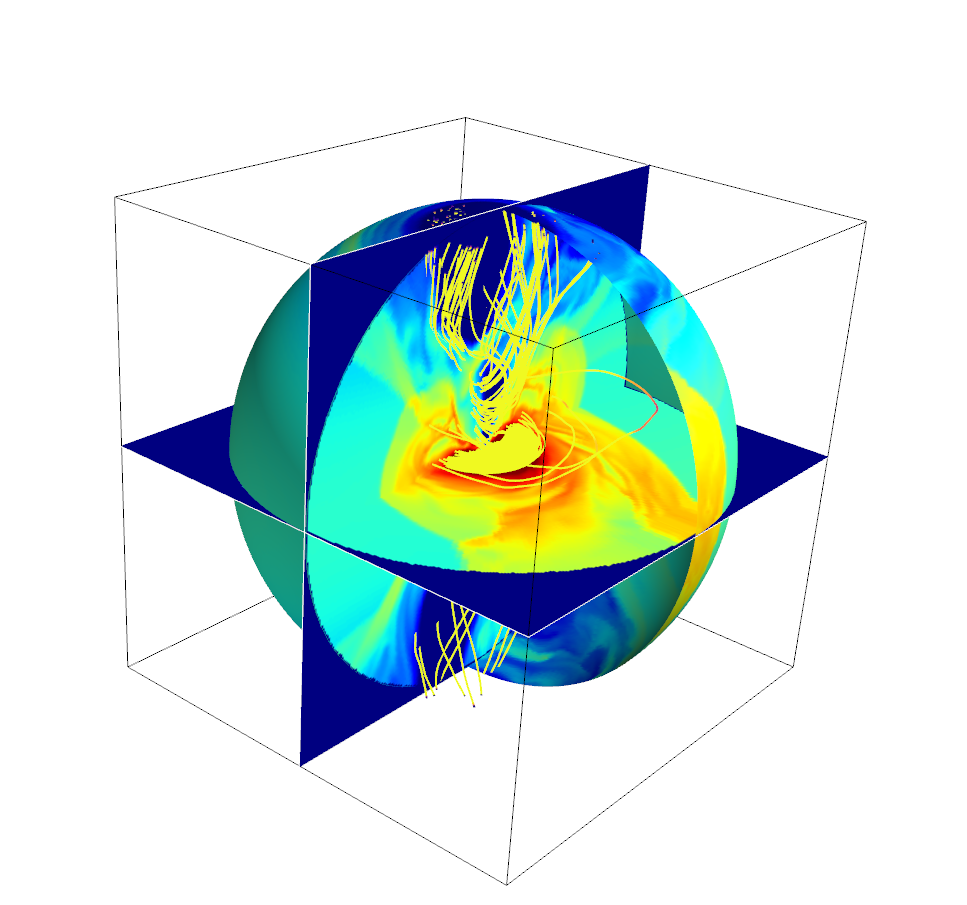}
        \caption{Gas streamlines in the vicinity of the perturber.}
        \label{fig:sub2}
    \end{subfigure}%
    \hfill

    \begin{subfigure}[b]{1.0\textwidth}
        \includegraphics[width=\linewidth]{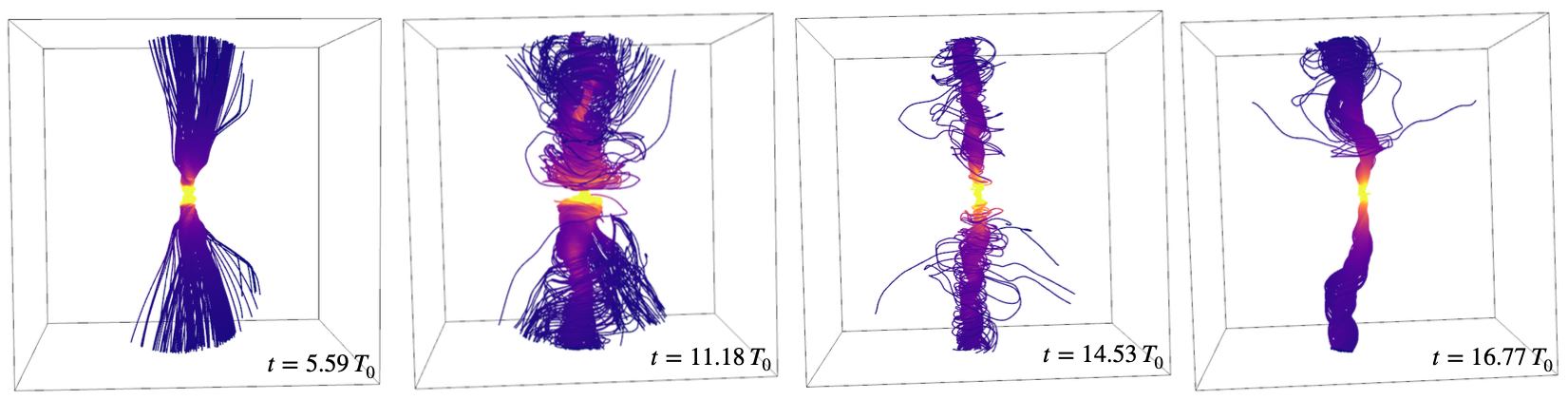}
        \caption{Temporal evolution of the rotational winding of the magnetic field lines.}
        \label{fig:sub3}
    \end{subfigure}%
    \hfill
    
    \caption{3D view of the density, the magnetic tornado and the bipolar jet within the perturber's Hill sphere.}
    \label{fig:multi}
\end{figure*}

Accretion onto a massive body embedded in a rotating disk arises in several astrophysical settings. Important examples include giant planets forming in protoplanetary disks
\citep[][]{Szulagyi2016,Tanigawa2012,Krapp2024,Sagynbayeva2025}
and compact objects embedded in active galactic nucleus (AGN) disks
\citep[][]{Dittmann2021,Chen_Ken2023,Jiao2025}.
Gas captured from the surrounding disk can circularize around the embedded object and form a mini-accretion disk
\citep[][]{Krapp2024}.

The structure of the mini-accretion disk is determined not only by the gravitational potential of the embedded body, but also by the properties of the surrounding disk, which sets the supply of mass, angular momentum, and magnetic flux. Hydrodynamical simulations with simplified thermodynamics provide a useful framework for characterizing the structure of the flow and the associated mass accretion rate
\citep[e.g.,][]{Sagynbayeva2025}.
This framework can then be extended to incorporate additional physical processes
\citep[e.g.,][]{2026arXiv260405020T}.

Magnetic fields are a natural next ingredient. They play a central role in accretion disk dynamics by regulating angular-momentum transport and mass accretion through turbulence driven by the magnetorotational instability
\citep[MRI;][]{BH1991,BH1998}
and through magnetized winds
\citep[][]{T2014}.
They may therefore also strongly influence the structure and dynamics of mini-accretion disks.

In this work, we present three-dimensional global simulations of a perturber embedded in a magnetized disk orbiting a central massive body. We investigate whether the gas captured within the Hill sphere forms a magnetically active mini-accretion disk, whether the magnetic field is amplified and reorganized by the local flow, and whether the resulting configuration can launch magnetized outflows. We adopt the ideal-MHD approximation as a baseline for isolating these processes before introducing the additional physics required for specific astrophysical applications.

The applicability of ideal MHD depends on the environment. In AGN disks, external ionizing sources, particularly nuclear X-rays, may maintain sufficient ionization for effective magnetic coupling out to large radii. Even gravitationally unstable regions may therefore approach the ideal-MHD regime in sufficiently luminous systems
\citep[][]{Menou2001}.
In protoplanetary disks, the ideal-MHD approximation is more restrictive. It may be justified only where the column density of the circumstellar and circumplanetary gas does not greatly exceed the characteristic cosmic-ray attenuation column,
$\chi_{\rm CR}\sim100\,{\rm g\,cm^{-2}}$
\citep[][]{UmebayashiNakano1981}.
This condition is unlikely to hold throughout circumplanetary environments
\citep[e.g.,][]{Fujii2014,Fujii2017}.
Recent polarization measurements of the HD~142527 disk, which suggest a plasma $\beta$ of order $200$
\citep[][]{Ohashi2025},
indicate that dynamically relevant magnetic fields can be present in planet-forming disks.

The present calculations should therefore be viewed as an ideal-MHD reference model. Future extensions will need to include non-ideal MHD effects, realistic heating and cooling, radiative transfer, and, in protoplanetary environments, dust dynamics
\citep[e.g.,][]{Krapp2024}.

\begin{figure*}
    \centering
    \includegraphics[width=1.0\linewidth]{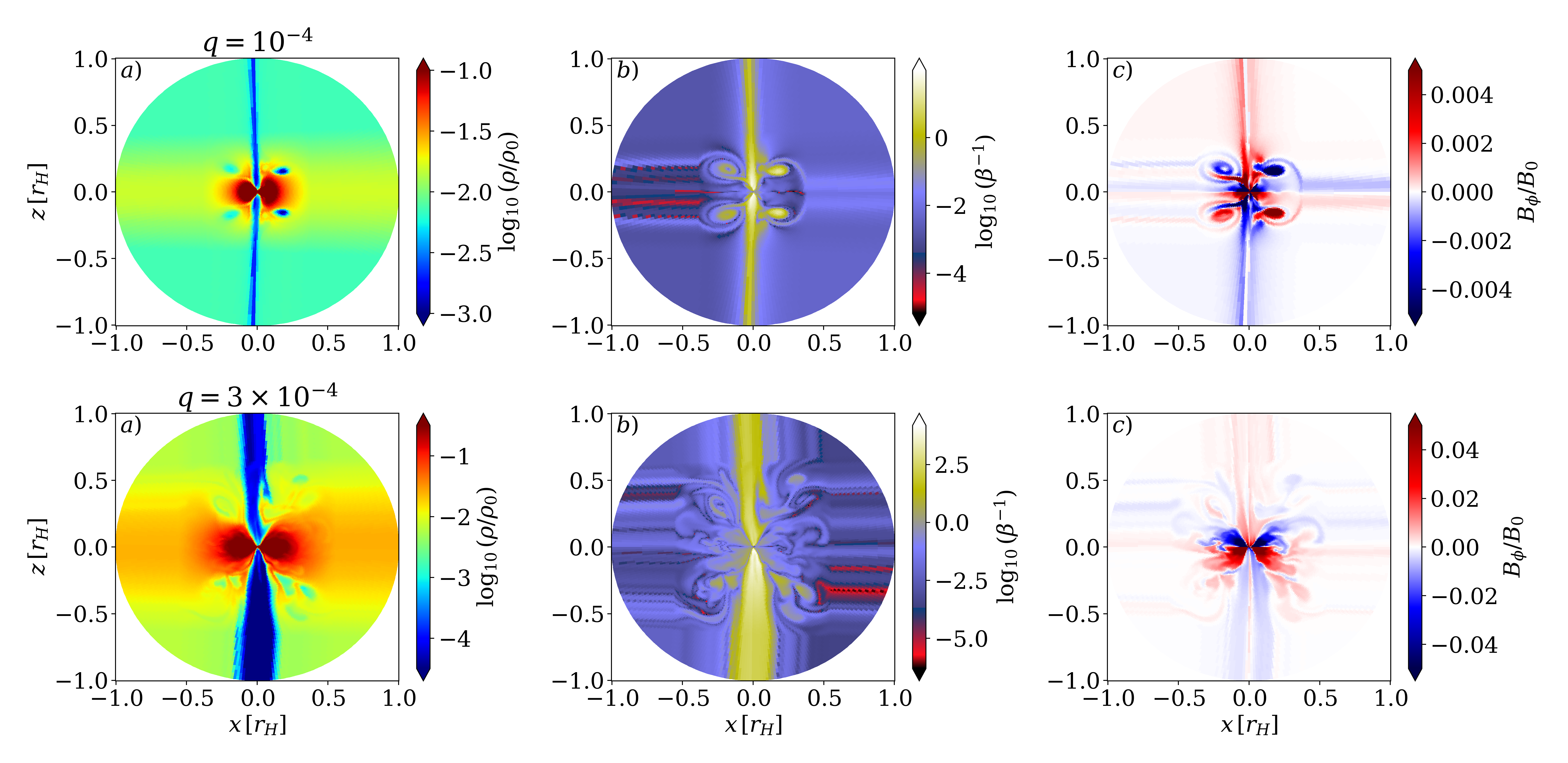}
    \caption{Vertical slices of the perturber's Hill sphere for $q=3\times10^{-4}$ and $q=10^{-4}$, at $t=33.54\,T_0$, in the $x-z$ plane, which is perpendicular to the direction of motion of the perturber. The various panels illustrate the gas density $\rho$, the inverse of the beta parameter $\beta^{-1}$, and the toroidal component of the magnetic field $B_{\phi}$.
    }
    \label{fig:q_lower}
\end{figure*}

\begin{figure}
    \centering
    \includegraphics[width=0.9\linewidth]{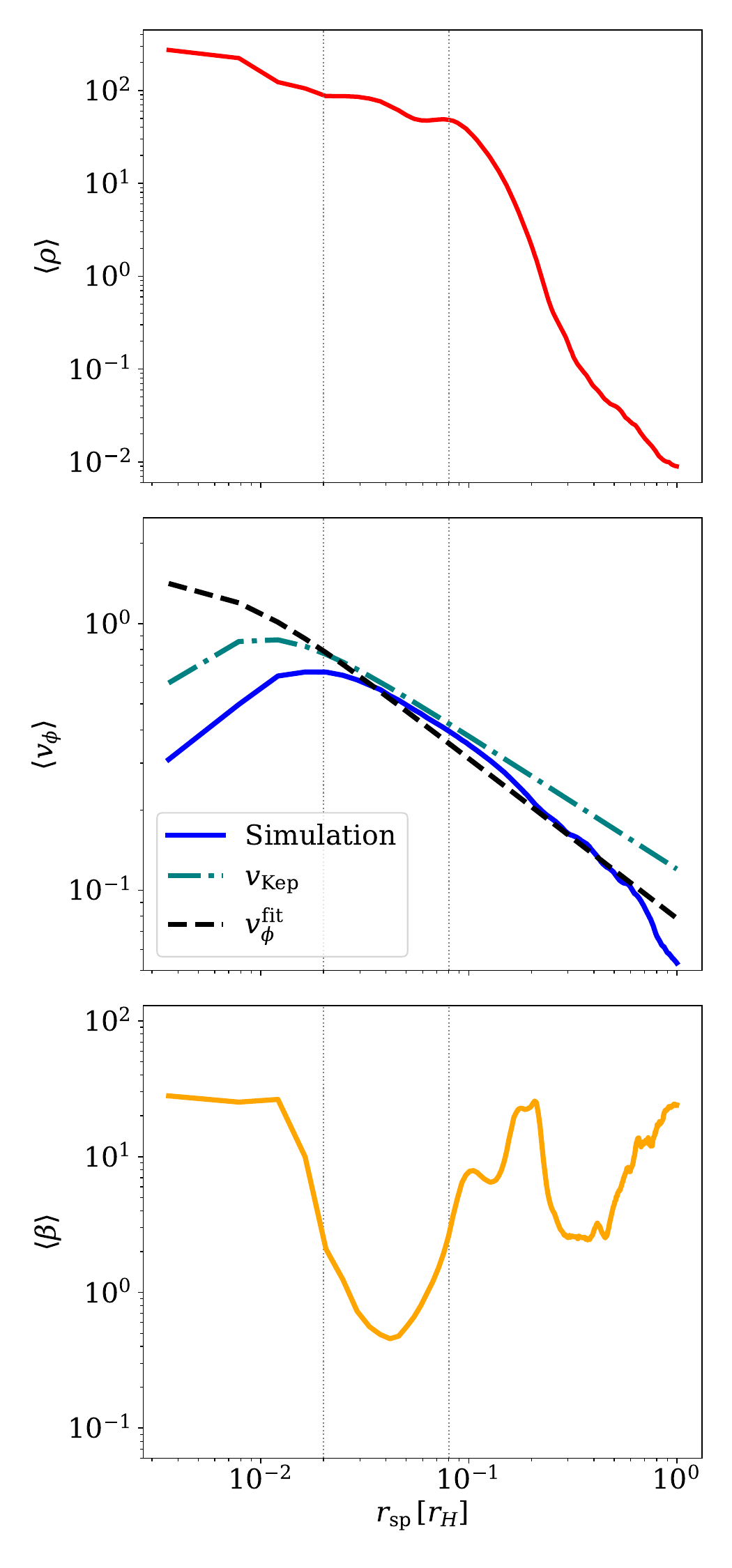}
    \caption{Radial profiles of the gas density, azimuthal velocity, and the plasma $\beta$ obtained from azimuthal averages in the mini-accretion disk midplane at $t=16.77\,T_\mathrm{0}$. In the middle panel,
    we also show the Keplerian velocity corresponding to the planetary gravitational potential (green curve),
    together with a fit ($v_\phi^{\rm fit}$)
    to the declining portion of the rotation velocity profile (black curve) (with $f_0=0.7$). The dotted vertical lines divide into three radial regions where different dynamical effects occur as consequence of the magnetic field (see text for details). The density and the azimuthal velocity are in units of $\rho_0=\Sigma_0/(\sqrt{2\pi}H_{\rm p})$ and $v_\mathrm{orb}= \sqrt{GM_{\rm c}/r_{\rm p}}$, respectively.
    }
    \label{fig:radial_prof}
\end{figure}

\begin{figure}
    \centering
\includegraphics[width=0.9\linewidth]{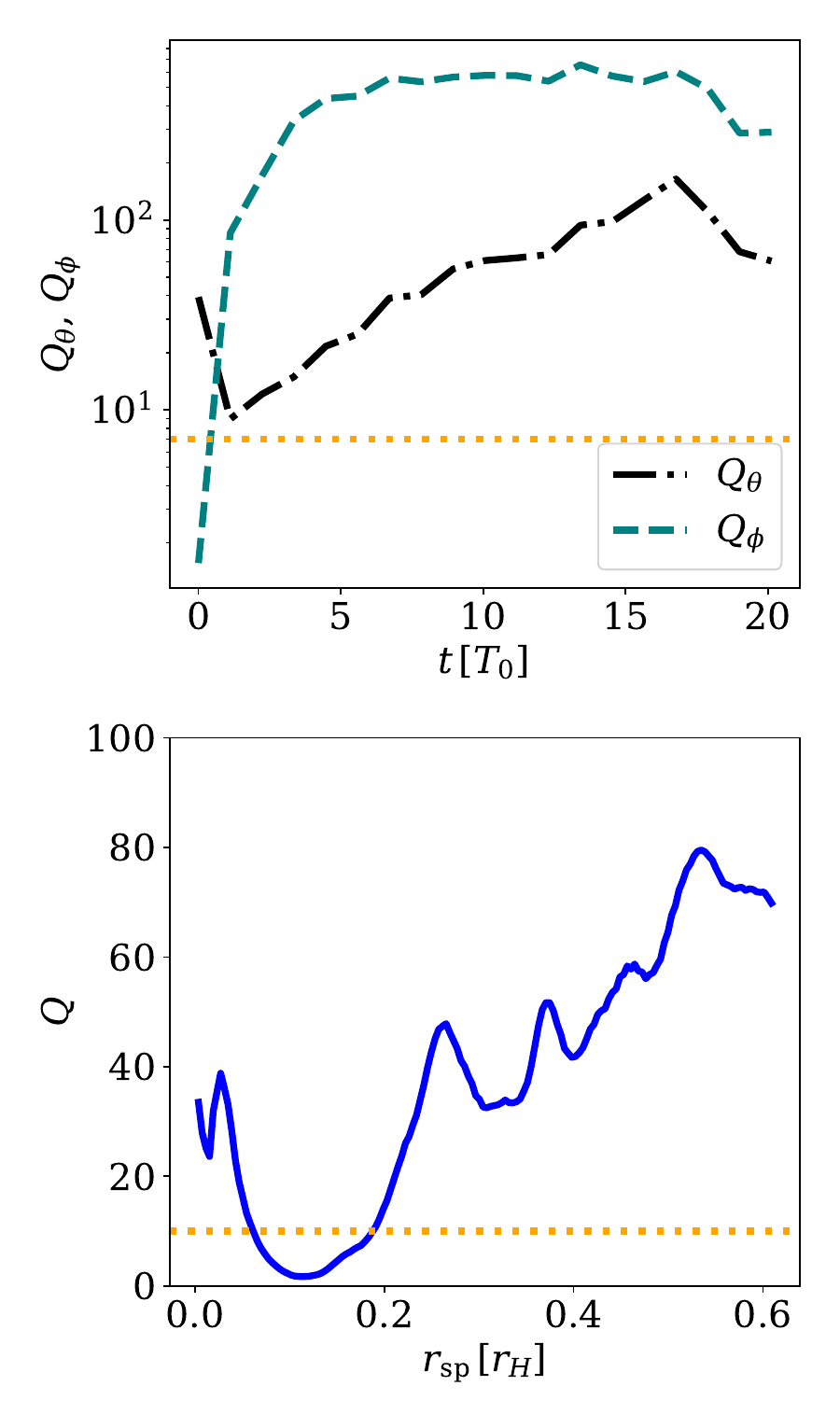}
    \caption{\textit{Top.} Quality factors defined in Eqs. (\ref{eq:Qz}) and (\ref{eq:Qp}), calculated in the frame of reference of the perturber. The dotted orange line represents the minimum value $Q_\theta$ and $Q_\phi$ that adequately solve $\lambda_\mathrm{MRI}$. \textit{Bottom.} New quality factor $Q$ defined in Eq. (\ref{eq:Q_R}).}
    \label{fig:Qua}
\end{figure}

\section{Physical model and numerical set-up}
\label{sec:equations}

We consider a stratified 3D locally isothermal accretion disk threaded by a uniform vertical magnetic field and hosting a perturber of mass $M_{\rm p}$ on a 
fixed circular orbit with radius $r_{\rm p}$.
The perturber mass ratio to that of the central object is $q\equiv M_p/M_c\in[10^{-4},10^{-3}]$. 
The gas dynamics is described by the ideal MHD equations:
\begin{equation}
\frac{\partial\rho}{\partial t}+\nabla\cdot(\rho \mathbf{v})=0,
\label{eq:gas_cont}
\end{equation}
\begin{equation}
\frac{\partial(\rho\mathbf{v})}{\partial t}+\nabla\cdot(\rho\mathbf{v}\mathbf{v}+p\mathbf{I})= -\rho\nabla\Phi+\mathbf{J}\times\mathbf{B},
\label{eq:gas_mom}
\end{equation}
\begin{equation}
\frac{\partial \mathbf{B}}{\partial t}=\nabla\times(\mathbf{v}\times\mathbf{B}),
    \label{eq:induc}
\end{equation}
where $\rho$ and $\mathbf{v}$ denote the gas density and the gas velocity, respectively, $\Phi$ is the gravitational potential, 
$p$ the gas pressure and $\mathbf{I}$ is the unit tensor. In Equations~(\ref{eq:gas_mom}) and~(\ref{eq:induc}), $\mathbf{J}=(\nabla\times\mathbf{B})/\mu_0$ and $\mathbf{B}$ are the current density and the magnetic field, respectively. The magnetic field obeys the solenoidal condition $\nabla\cdot\mathbf{B}=0$. For the gas pressure, we consider the equation of state
\begin{equation}
 p=c_s^2\,\rho, \label{eq:pressure}
\end{equation}
where $c_s$ is the isothermal sound speed. 

The gravitational potential $\Phi$ acting on the fluid includes contributions from the central object and the perturber, as well as the so-called indirect term arising from the reflex motion of the central object \citep[][]{MD1999,Muller2012,Crida2025}. More
specifically,
\begin{equation}
\Phi=\Phi_{\rm c}+\Phi_{\rm p},
 \label{eq:potential}
\end{equation}
where
\begin{equation}
\Phi_{\rm c}=-\frac{GM_{\rm c}}{r},
 \label{eq:Star_potential}
\end{equation}
and
\begin{equation}
\Phi_{\rm p}=-\frac{GM_{\rm p}}{\sqrt{|\mathbf{r}-\mathbf{r}_{\rm p}|^2+\epsilon^2}}+\frac{GM_{\rm p} r\cos\phi\sin\theta}{r_{\rm p}^2},
 \label{eq:Planet_potential}
\end{equation}
where $\mathbf{r}_{\rm p}$ is
the perturber's position. Its gravitational potential is softened with a length scale $\epsilon = 0.0072\,r_{\rm H}$, similar to the cell size of the local grid (see below). Throughout this paper, $r_{\rm H} = r_{\rm p} (q/3)^{1/3}$ denotes the 
perturber's Hill radius. 

The initial gas density is given by 
\begin{equation}
 \rho\left(r,\theta\right)=\rho_\mathrm{eq}\left(r\right)\,(\sin\theta)^{-\sigma-\xi+h^{-2}},\label{eq:rhog}
\end{equation}
with 
\begin{equation}
 \rho_\mathrm{eq}\left(r\right)=\frac{\Sigma_0}{\sqrt{2\pi}h r_{\rm p}}\left(\frac{r}{r_{\rm p}}\right)^{-\xi},   \label{eq:rhoeq}
\end{equation}
while the initial velocity components\footnote{For the purpose of highlighting in which reference system the velocity and magnetic field fields are shown, we add the subscript 'c' to each of the components, which refers to the fact that they are calculated in the frame fixed on the central object.} are $v_{r,\rm c}=v_{\theta, \rm c}=0$ and 
\begin{equation}
  v_{\phi,\rm c}=\sqrt{\frac{GM_{\rm c}}{r\sin{\theta}}-\xi c_s^2} \label{eq:vphi}
\end{equation}
\citep[see Appendix A of][]{MB2016}.
We adopt code units such that $M_{\rm c}=1$, $r_{\rm p}=1$ and 
the surface density at $r_{\rm p}$ is $\Sigma_0=2\times10^{-3}/\pi$. We set the values $\sigma=1$, $\xi=1.5$
and $h=0.05$.

For the magnetic field, we adopt an initially uniform vertical configuration with 
a plasma parameter 
\begin{equation}
 \beta\equiv 2\mu_0c_s^2\rho/B^2=875,  \label{eq:pla}
\end{equation}
evaluated at the disk midplane at $r_{\rm p}$. For a protoplanetary disk with $M_{\rm c}=1M_{\odot}$ 
and $r_{\rm p}=5.2$ au, this corresponds to a vertical magnetic field strength of $B_z=54\,\mathrm{mG}$ \citep[see also][]{Gressel2013}.

We use the public code \textsc{Fargo3D} \citep[][]{BllM2016} and solve the MHD equations in spherical coordinates centered on the central object, in a frame co-rotating with the perturber. Nevertheless,  we primarily show the results in interpolated\footnote{We use a nearest interpolation using $256^3$  evenly spaced grid cells within a sphere of radius $r_\mathrm{H}$.} spherical coordinates centered on the perturber $(r_{\rm sp},\theta_\mathrm{sp},\phi_\mathrm{sp})$. 

The numerical domain extends from $0.48 r_{\rm p}$ to $2.08 r_{\rm p}$ in the $r$-direction, from $-\pi$ to $\pi$ in the $\phi$-direction, and between $\pi/2-2h$ and $\pi/2+2h$ in colatitude. To increase resolution inside the Hill sphere, where the mini-accretion disk forms, we use mesh density functions\footnote{To make use of the mesh density functions in the azimuthal direction, we consider the standard transport in our simulations. In the radial and azimuthal directions, we use the coefficients $a_r=a_\phi=0.2618$, $b_r=b_\phi=0.3141$, $c_r=15.0$ and $c_\phi=1.5$, respectively. For colatitude, we follow the approach described in \citet{Chametla2025}, which leads to cubic grid cells.}
\citep[see][for details]{Bll2023}. 
The mesh is refined around the perturber and extends beyond the Hill sphere, maintaining a constant grid spacing of $5\times10^{-4}r_{\rm p}$ in each direction, which we achieve by including $(N_r,N_\theta,N_\phi)=(1864,256,1864)$ grid zones throughout the computational domain. We include magnetic resistivity buffer zones only near the boundaries in the radial direction of the global disk, which cover the same computational domain as the damping zones of the hydrodynamic variables \citep[][]{deVal2006}.  

We use the orbital period, $T_0$, of a fluid element within the mini-accretion disk located at radius of $0.5\,r_{\rm H}$  
as unit of time. This corresponds to approximately $0.2$ orbital periods of the perturber. The duration of the simulations was limited by the computational cost of running
them for longer. Nevertheless,
we evolve the global disk at least for $9T_{0}$ ($1.8$ perturber's orbits), which is sufficient time for the overall structure of the mini-accretion disk to be well defined \citep[see, for instance,][]{Fung2019}. 
However, it is likely that the viscous timescale of the mini-accretion disk,
$\tau_{\nu}$, exceeds our simulation runtime.
The value of $\tau_{\nu}$ depends on its effective turbulent viscosity, 
which is determined by the turbulent state of the mini-accretion disk and is therefore not known a priori, but is instead  an
outcome of the simulations. Assuming that the disk
has reached a steady state, then $\tau_{\nu} \equiv (0.5r_{\rm H})^{2}/\nu \simeq 0.5 \alpha^{-1} h^{-2} q^{2/3} T_{0}$,
where $\alpha$ is the Shakura-Sunyaev viscosity parameter\footnote{We have used that $\nu=\alpha h^{2} v_{\rm orb}^{2} T_{0}/(2\pi)$ with $v_{\rm orb}\equiv \sqrt{GM_{\rm c}/r_{\rm p}}$ and $T_{0}\simeq (2\pi/5)(r_{\rm p}/v_{\rm orb})$.
Therefore, $\tau_{\nu}= \lambda \alpha^{-1} h^{-2} q^{2/3} T_{0}$, with $\lambda\simeq (25/4)(1/3)^{2/3}/(2\pi)\simeq 0.5$.}.
For $\alpha\simeq 0.03-0.1$, $h=0.05$ and $q=10^{-3}$,
this gives $\tau_{\nu}\simeq (20-60)T_{0}$.
As shown in Section \ref{sec:discussion}, however, the mini-accretion disk remains
in an early dynamical stage, so its evolution cannot be described
by a quasi-steady viscous inflow.

\begin{figure}
    \centering
    \includegraphics[width=\linewidth]{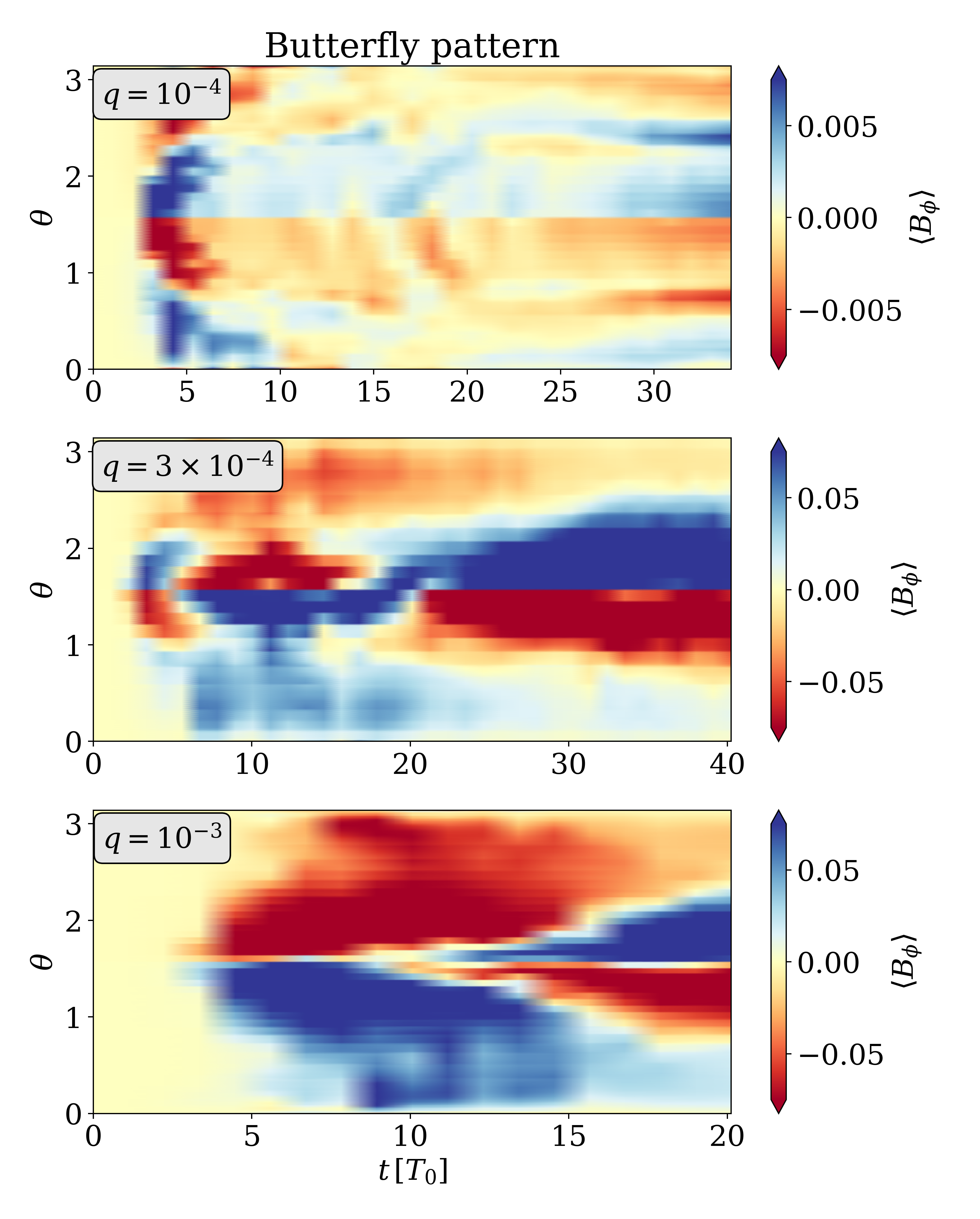}
    \caption{Space-time diagram of the azimuthally averaged $B_{\phi}$, denoted by $\langle B_\phi \rangle$, as a function of colatitude (normalized by $B_0$) for $q\in[10^{-4},10^{-3}]$.}
    \label{fig:butt}
\end{figure}

\section{The magnetic tornado}
\label{sec:results}

We first focus on the case where the perturber has $q=10^{-3}$. Fig. \ref{fig:global} shows a vertical cut along the radial direction of the gas density and azimuthal component of the magnetic field from the system of reference centered in the central object. The mini accretion disk can be seen edge-on. The accreting flow onto the perturber has differential rotation that induces the azimuthal magnetic field.

Fig. \ref{fig:fields} presents vertical slices in the same plane but a zoom of the perturber's Hill sphere at $t=16.77\,T_0$. The gas density map shows the formation of a thick mini-accretion disk around the perturber which, together with the maps of the poloidal-to-toroidal ratios for the components of the velocity and magnetic fields, indicates a turbulent state. In addition, low-density bipolar jets emerge along the $z$-axis. The footpoints of the magnetic field lines entrained in the bipolar jets are on the inner part of the mini-accretion disk within $r_{\rm sp}\leq r_\mathrm{bj}=0.072\,r_{\rm H}$. Most of the outflow activity is contained in a narrow cone with an opening angle of $\sim 25^{\rm o}$ from the vertical axis. In this region, strong magnetically-dominated flows develop, as indicated by $\beta^{-1}>1$, panel $b)$ in Fig. \ref{fig:fields}.  Within the jet-launching region, we observe: 
{\it i}) the development of a strong toroidal component of the magnetic field, panel $c)$  in Fig. \ref{fig:fields}, 
{\it ii}) the poloidal components of both the velocity and the magnetic field can dominate,
and {\it iii}) a low poloidal Alfvén Mach number $\mu_\mathrm{A,p}=v_{\rm p}\sqrt{4\pi\rho}/B_{\rm p}$. These three features are characteristic signatures of magnetically driven jet formation in accretion disks \citep[][]{Vetter2025}. We recall that for the magnetocentrifugal mechanism to operate, it is sufficient that the poloidal component of the magnetic field be of the order of the toroidal component, that is $B_{\rm p}\sim B_{\phi}$ \citep[e.g.,][]{C1995}. This condition is indeed satisfied in the region where the jet is launched, see panel $e)$ in Fig. \ref{fig:fields}. Therefore, our simulation suggests that the bipolar jets are driven by the strong toroidal magnetic field component $B_\phi$ in the inner region  \citep[plasma-gun mechanism:][]{C1995} with a contribution from magnetocentrifugal acceleration as suggested by \citet{BP1982}.

To test this hypothesis, we have compared the vertical component of the acceleration due to magnetic pressure 
\begin{equation}
    a_{\mathrm{mag},z} = -\frac{1}{\rho}\left(\frac{\partial P_{\mathrm{mag}}}{\partial r_\mathrm{sp}}\cos\theta_\mathrm{sp}-\frac{1}{r_\mathrm{sp}}\frac{\partial P_{\mathrm{mag}}}{\partial \theta_\mathrm{sp}}\sin\theta_\mathrm{sp}\right),
\end{equation}
and the vertical component of the gravity acceleration from the perturber
\begin{equation}
    a_{\mathrm{grav},z}=-\frac{GM_p}{r_{\mathrm{sp}}^2}\cos\theta_\mathrm{sp},
\end{equation}
where $P_{\mathrm{mag}}=(B_{r, \rm c}^2+B_{\theta, \rm c}^2+B_{\phi, \rm c}^2)/(2\mu_0)$. We show the results of this comparison in Fig. \ref{fig:gradPm}. We can see that the acceleration ratio exceeds unity ($a_{\mathrm{mag},z}/a_{\mathrm{grav},z}> 1$) as one approaches the polar regions. This demonstrates that the perturber's gravitational force is unable to confine the gas at the poles, where the magnetic pressure gradient becomes the primary dynamic engine, accelerating the flow vertically and sustaining the structure of the bipolar jets.

The magnetic-to-gravitational acceleration ratio is not symmetric with respect to the midplane at high latitudes, implying an asymmetry in the jet power. This asymmetry can be observed also in the gas density in Figs. \ref{fig:fields} and \ref{fig:q_lower}. This behavior is consistent with the findings of \citet{Fendt2013}, for a circumstellar disk, where the disk sustain long-term ejection asymmetries, a phenomenon that has also been observationally mapped in bipolar flows \citep{Podio2016,Dutta2025}.

In addition, we have analyzed the structure of the magnetic field lines and gas streamlines in the low-density region along the rotation axis of the mini-accretion disk (Fig. \ref{fig:sub1} and \ref{fig:sub2}, respectively). A tornado-like structure is visible in the magnetic field lines. 
In the midplane, we identify a mini-accretion disk extending to $r_{\rm out} \sim 0.6\,r_{\rm H}$ (see also  Fig. \ref{fig:fields}a) in which the gas velocity is mostly azimuthal except in the innermost region, where an outflow associated with the magnetic tornado is evident.
We find a characteristic pattern of twisting of the magnetic field lines generated by the rotation of the gas inside the mini-accretion disk, which is observed in the temporal evolution shown in Fig. \ref{fig:sub3}. At early times ($t=5.59\,T_0$), the concentration and twisting of the magnetic field lines near the poles of the perturber is evident. As time elapses ($t\geq11.18\,T_0$), the twisted field lines rise to higher altitudes, eventually reaching the boundaries of the Hill sphere. Finally, at $t=16.77\,T_0$, coherent columns of twisted magnetic field lines can be observed at the polar regions around the perturber.

On the other hand, in Fig. \ref{fig:q_lower} we show vertical slices (in the ($x,z$) plane within the perturber's Hill sphere) of the gas density, beta parameter and the toroidal component of the magnetic field for the models with $q=10^{-4}$ and $q=3\times10^{-4}$ at $t=33.54\,T_0$. In both cases, the formation of bipolar jets in the gas density is clearly visible. In this jet formation region, the dynamics are again dominated by the magnetic field since $\beta^{-1}>1$, as well as by the development of the toroidal magnetic field component.  

Our results generalize previous findings, focused on mass ratios $q=(3-4)\times 10^{-4}$ \citep[][]{Machida2006,Gressel2013}, to lower-mass perturbers and demonstrate the robustness of the jet-launching mechanism in magnetized mini-accretion disks. It is worth noting that this is the first time bipolar jet formation has been observed in perturbers with a mass ratio of $q=10^{-4}$. 
In terms of $v_{\rm orb}\equiv \sqrt{GM_{\rm c}/r_{\rm p}}$, the jet velocity increases from $0.87v_{\rm orb}$ to about $2.3v_{\rm orb}$ after $10\,T_{0}$, and remains near this value thereafter, 
thereby exceeding not only the escape speed from the perturber ($0.17v_{\rm orb}$  at the edge of the Hill sphere) but also the escape velocity from the 
central object ($1.4v_{\rm orb}$). 

If these results are rescaled to a Jupiter-mass planet at $r_{\rm p}=5.2$ au around a solar-like star,
this gas would eventually be ejected into the interstellar medium at velocities of $\sim 30\,\mathrm{km\,s^{-1}}$. Bipolar jets with these characteristics can be potentially detectable. 
The presence of large-scale outflows motivate the use of observational techniques (e.g., employing sulfur monoxide SO shock tracer or kinematics of Na D) to detect outflows in accretion disks where massive perturbers are thought to be forming \citep[][and references therein]{Dutrey2024,Zakamska2025}. 
In fact, jets with velocities in that range have already been detected \citep[][]{Zakamska2025}.

The outflows we observe could also have implications for large-scale dust dynamics across the gaps formed by the planets. Expanding our framework to incorporate further physics, such as non-ideal MHD \citep{G2015,G2020}
multi-species dust dynamics \citep{Bll2019} and more realistic thermodynamics \citep{Krapp2024}, will make it possible to study whether large-scale planetary outflows could modify dust filtering across gaps formed by Jupiter-like planets \citep{2018ApJ...854..153W}. If persistent, these outflows could have potential implications for the dichotomy observed in the sizes of calcium–aluminum-rich inclusions (CAIs) seen in chondrites \citep{2019AJ....158...55H}.

\section{The turbulent mini-accretion disk}
\label{sec:discussion}

In Fig. \ref{fig:radial_prof} we show the radial profiles of the gas density $\rho$, azimuthal velocity $v_\phi$, and the plasma $\beta$, averaged azimuthally in the mini-accretion disk midplane for the model with $q=10^{-3}$, calculated at $t=16.77T_0$. In addition to the Keplerian velocity profile from the gravitational potential of the planet
\begin{equation}
    v_{\rm Kep}=\frac{\sqrt{GM_{\rm p}}\,r_{\rm sp}}{\left(r_{\rm sp}^2+\epsilon^2\right)^{3/4}}
\end{equation}
we include a fit to the declining portion of the rotation velocity profile, given as
\begin{equation}
    v_\phi^{\rm fit}=f_0\, v_{\rm Kep}(0.5r_H)\left(\frac{\sqrt{(0.5r_H)^2+\epsilon^2}}{\sqrt{r_{\rm sp}^2+\epsilon^2}}\right)^{3/5}
\end{equation}
with $f_0=0.7$. We identify three distinct disk regions: 

\begin{itemize}
    \item the innermost region ($r_\mathrm{sp} < 0.02\,r_{\rm H})$, characterized by the
    highest volume densities, a radially increasing subKeplerian
    azimuthal velocity and a 
    strong predominantly poloidal magnetic field.
    \item an intermediate region ($0.02\,r_{\rm H} < r_\mathrm{sp} < 0.08\,r_{\rm H}$), where the gas density exhibits a smooth decrease, the azimuthal velocity is almost Keplerian, and the pressure is dominated by the toroidal component of the magnetic field, which is main region where the bipolar outflows originate, and
    \item an outer disk region ($r_\mathrm{sp}>0.08\,r_{\rm H}$), where density decreases monotonically, approximately as $r_\mathrm{sp}^{-7/2}$, the azimuthal velocity decays slightly faster than a Keplerian profile, with $v_\phi \propto \left(1.0/\sqrt{r_\mathrm{sp}^2 + \epsilon^2}\right)^{\frac{3}{5}}$, and the plasma $\beta$ is larger than one.
\end{itemize}

\begin{figure}
    \centering
    \includegraphics[width=\linewidth]{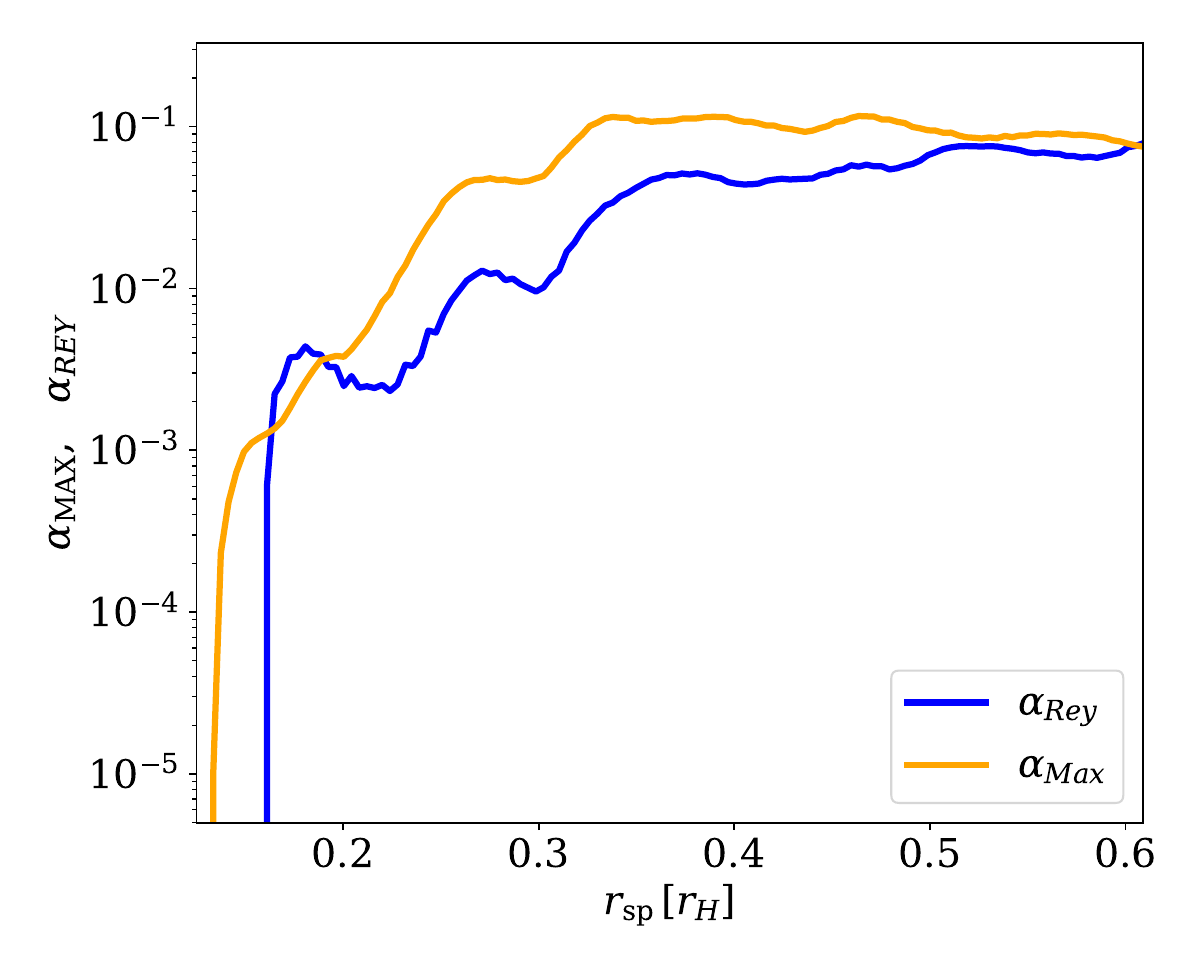}
    \caption{Radial profiles of $\alpha_\mathrm{MAX}$ and $\alpha_\mathrm{REY}$  (azimuthally and temporally averaged between 16.77 to 20.12 $T_0$) at the mini-accretion disk midplane for $q=10^{-3}$.}
    \label{fig:alpha}
\end{figure}

These results indicate that the conditions required for the onset of the MRI, namely an outwardly decreasing angular velocity and a relatively weak
magnetic field ($\beta\gtrsim 1$), are satisfied throughout
most of the mini-accretion disk.
To show that the MRI is indeed active and well resolved, we compute two figures of merit that have been proposed to this end. It has been suggested that properly resolving the MRI with a grid size $\Delta \theta$, $\Delta \phi$ requires that the quality factors $Q_\theta$ and $Q_\phi$ defined as
\begin{equation}
    Q_\theta\equiv\frac{\lambda_\mathrm{MRI}}{R\Delta \theta}=2\pi\sqrt{\frac{16}{15}}\frac{v_{A,\theta}}{\Omega R\Delta \theta} \,
    \label{eq:Qz}
\end{equation}
and
\begin{equation}
    Q_\phi\equiv\frac{\lambda_\mathrm{MRI}}{R\Delta \phi}=2\pi\sqrt{\frac{16}{15}}\frac{v_{A,\phi}}{\Omega R\Delta \phi} \,
    \label{eq:Qp}
\end{equation}
\citep[][]{Haw2011}, respectively, must be $\geq 7$ \citep{Sor2012}. 
Here, $\lambda_\mathrm{MRI}$ stands for the wavelength of the fastest-growing MRI-unstable mode, $v_{A,\theta}$ and $v_{A,\phi}$ are the polar and azimuthal Alfv\'en speed in that directions, $R=r_\mathrm{sp}\sin\theta_\mathrm{sp}$ is the cylindrical radius, and $\Omega$ is the local gas angular frequency. To assess whether this condition is satisfied, Fig. \ref{fig:Qua} shows the temporal evolution of $Q_\theta$ and $Q_\phi$ (radially, vertically, and azimuthally averaged). It is clear that for $t\geq3T_0$ the condition $Q_\theta\geq 7$ is fully satisfied. Recently, an alternative quality factor for global simulations has been introduced by \citet{JL2025}:
\begin{equation}
Q(R)=\frac{\sqrt{3}B_\mathrm{net}}{\sqrt{\pi\rho_\mathrm{mid}}\Omega R\Delta \theta} \,,
    \label{eq:Q_R}
\end{equation}
where $B_\mathrm{net}$ is the vertical component of the magnetic field (averaged vertically and azimuthally) and  $\rho_\mathrm{mid}$ is the gas density at the midplane. If we adopt the threshold criterion for the development of MRI suggested by \citet{JL2025} ($\sim80\%$ of the injected energy), $Q$ must be $\geq 10$. This alternative condition is also satisfied in almost the entire mini-accretion disk at $t=16.7\,T_0$ as can be seen in the lower panel of Fig. \ref{fig:Qua}. Note that our results are not expected to be sensitive to azimuthal resolution, unless the cells are substantially more elongated in the azimuthal direction than in the vertical direction. For comparable physical grid scales in the two directions (as is the case here), shear tends to stretch turbulent structures preferentially in the toroidal direction. As a result, the characteristic azimuthal coherence lengths are expected to be larger than the corresponding vertical scales. Structures that are adequately resolved vertically should therefore also be resolved in the azimuthal direction (see $Q_\phi$ in Fig. \ref{fig:Qua}).

\begin{figure}
    \centering
    \includegraphics[width=0.7\linewidth]{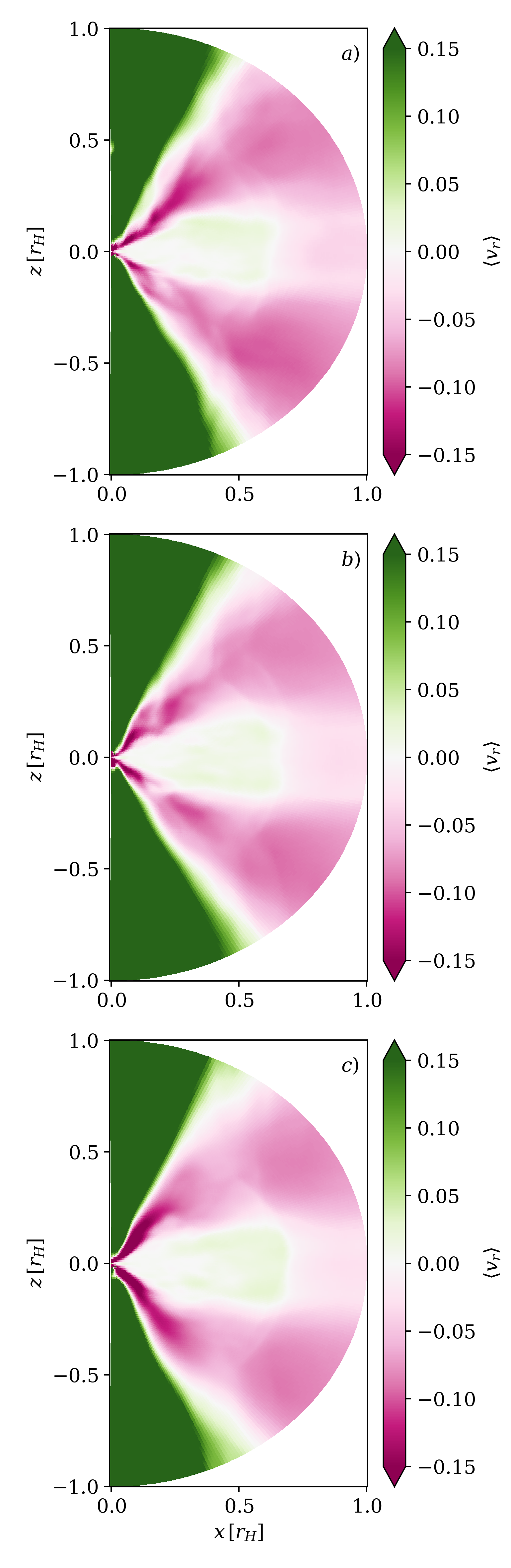}
    \caption{Azimuthally averaged radial velocity at three different times: at $t=16.77T_0$ (panel a), $t=17.88T_0$ (panel b) and $t=19.00T_0$ (panel c), for a model with $q=10^{-3}$.}
    \label{fig:vrad}
\end{figure}

The dynamics of the large-scale toroidal field in the mini-accretion disk shows organized polarity reversals and coherent space-time structure (see Fig. \ref{fig:butt}). The mini-accretion disk therefore amplifies and organizes the magnetic field, producing large-scale toroidal-field patterns reminiscent of the butterfly diagrams observed in stratified MRI turbulence \citep[e.g.,][]{Brandenburg1995,Pessah2010,2010MNRAS.405...41G,2015ApJ...810...59G}. These features are consistent with MRI-driven dynamo activity, although establishing this would require a full mean-field diagnosis, including the decomposition of the mean induction equation, the measurement of the turbulent electromotive force, and, ideally, the determination of the relevant dynamo coefficients involving test-fields \citep[as in, e.g.,][]{2015ApJ...810...59G}. A full mean-field analysis of the turbulent mini-accretion disk dynamics is, however, beyond the scope of the present paper.

For the smallest value of $q=10^{-4}$, the butterfly pattern appears more blurred. This is explained by the fact that as $q$ decreases, the size of the perturber's Hill radius also decreases, resulting in lower resolution. Additionally, we found that the disk thickness increases slightly more than in the other two cases, which also contributes to a more disordered butterfly pattern \citep[][]{Hogg2018}.

In Fig. \ref{fig:alpha} we show the radial profiles of turbulent Maxwell and Reynolds stresses, $\alpha_\mathrm{MAX}$ and $\alpha_\mathrm{REY}$ \citep[azimuthally and temporally averaged, see][]{Chametla2024} for $q=10^{-3}$ model, which describe the transfer of angular momentum through the mini-accretion disk. In the MRI-unstable outer disk region identified earlier, these profiles, with $\alpha_{\rm MAX}$ generally greater than $\alpha_{\rm REY}$, are characterized by $\alpha \sim 10^{-2} - 10^{-1}$, conditions that are not uncommon for MRI-driven turbulence in simpler settings \citep[see][and references therein]{2007ApJ...668L..51P}. 

Fig. \ref{fig:vrad} shows the radial velocity, averaged over azimuth, at three different times for $q=10^{-3}$. In analogy to the work of \citet{Mishra2020} for magnetized accretion disks, we identify three regions based on the aperture parameter $|z/R|$, where $R$ is the cylindrical radius: 1) the polar cone defined by $|z/R|>2$ where the flow moves outwards and form the bipolar jet; 2) the intermediate region, 
$0.2\leq |z/R|\leq2.0$, where the flow is predominantly accreting ($\langle v_r\rangle <0$); and 3) the
equatorial region $|z/R|<0.2$, where the mini-accretion disk is dense and $\langle v_r\rangle$ is small but positive.

\section{Future directions}

This work presents global ideal-MHD simulations of an embedded perturber in which the bound gas forms a turbulent, magnetized mini-accretion disk and launches a collimated outflow on scales comparable to the Hill radius. Under the assumptions adopted here, the mini-disk amplifies magnetic fields via a dynamo, sustains turbulent Maxwell stresses, and enters an outflow-launching regime without requiring a specially imposed local field geometry. The present models should therefore be regarded as a baseline: they isolate the role of ideal-MHD turbulence and magnetic-field amplification, while leaving the additional physical ingredients required for specific astrophysical applications to future work.

One natural extension is to compact objects embedded in AGN disks, where stellar-mass black holes and black-hole binaries may interact with the surrounding gas through accretion, migration, alignment, and binary hardening \citep{2017ApJ...835..165B,2020ApJ...898...25T,2025MNRAS.544.4576R,2026MNRAS.547ag427F}. Recent three-dimensional MHD simulations of embedded circum-single disks and binary black holes in AGN disks highlight the importance of magnetic fields in this problem \citep{2024arXiv240905614M,2025A&A...703A.304J}. The simulations presented here identify local MHD processes---magnetic-field amplification, turbulent angular-momentum transport, and outflow launching---that future simulations of black holes embedded in AGN disks will need to capture. Such extensions may help determine whether gas dynamics in AGN disks leaves observable imprints on compact-binary populations and on possible electromagnetic counterparts to gravitational-wave events \citep{2023ApJ...950...13T,2024ApJ...966...21T,2025PhRvD.111h3020R,2025PhRvD.111h3033M}.

At lower mass ratios, the same numerical approach can be applied to giant planets and their circumplanetary environments. In protoplanetary disks, however, the ideal-MHD approximation is only a first step. Non-ideal MHD effects, realistic thermodynamics, radiative transfer, and dust dynamics are likely to influence the coupling of the gas to the magnetic field, the thermal structure of the circumplanetary flow, and the delivery of solids to the planet's environment \citep{Ayliffe2009,2018ApJ...865..105K,Krapp2024}. Including these ingredients will be necessary to determine whether magnetized outflows persist under planet-forming conditions and whether they affect accretion onto forming planets, angular-momentum transport in circumplanetary disks, dust filtering across planetary gaps, or satellite formation \citep{2018ApJ...854..153W,2019AJ....158...55H,BatM2020ApJ}.

\begin{acknowledgements}
The authors thank the referee for the constructive reports and for the helpful comments and suggestions. 
The work of R.O.C was supported by the Czech Science Foundation (grant 25-16507S).  The work presented here is supported by the Carlsberg Foundation, grant CF25-1297. Computational resources were available thanks to the Ministry of Education, Youth and Sports of the Czech Republic through the e-INFRA CZ (ID:90254).
\end{acknowledgements}

\bibliographystyle{aa} 
\bibliography{example} %

\end{document}